\documentclass[conference]{IEEEtran}
\IEEEoverridecommandlockouts
\usepackage{cite}
\usepackage{amsmath,amssymb,amsfonts}
\usepackage{algorithmic}
\usepackage{graphicx}
\usepackage{verbatim}
\usepackage{textcomp}
\usepackage{xcolor}
\def\BibTeX{{\rm B\kern-.05em{\sc i\kern-.025em b}\kern-.08em
    T\kern-.1667em\lower.7ex\hbox{E}\kern-.125emX}}
\usepackage{caption}
\usepackage{subcaption}
\usepackage{multirow}
\usepackage{url}
\usepackage{pifont}
\usepackage[most]{tcolorbox} 
\usepackage[hidelinks]{hyperref}

\definecolor{brown(web)}{rgb}{0.65, 0.16, 0.16}

\newtcolorbox[auto counter]{tkx}[2][]{%
    enhanced, breakable, center title,
    colframe = #2!45,
    colback  = #2!10,
    colbacktitle=#2!20,
    left=-0.5pt,
    right=-0.5pt,
    bottom=-2pt,
    top=-0.25pt,
    #1% 
}
\newcounter{obs}
\newcommand\observation[1]{
\refstepcounter{obs}
\begin{tkx}{brown(web)}
\noindent\textbf{Finding~\theobs.} #1
\end{tkx}
}

\newcommand{\dingOne}{\ding{182}}
\newcommand{\dingTwo}{\ding{183}}
\newcommand{\dingThree}{\ding{184}}
\newcommand{\dingFour}{\ding{185}}

\begin{document}

\title{Investigating Memory Failure Prediction Across CPU Architectures\vspace{-0.2cm}}

\author{

\IEEEauthorblockN{Qiao Yu\IEEEauthorrefmark{1}\IEEEauthorrefmark{2},
Wengui Zhang\IEEEauthorrefmark{3},
Min Zhou\IEEEauthorrefmark{3}\IEEEauthorrefmark{5}\thanks{\IEEEauthorrefmark{5}Corresponding author: zhoumin27@huawei.com},
Jialiang Yu\IEEEauthorrefmark{3},
Zhenli Sheng\IEEEauthorrefmark{3},
Jasmin Bogatinovski\IEEEauthorrefmark{2}\\
Jorge Cardoso\IEEEauthorrefmark{1}\IEEEauthorrefmark{4} and
Odej Kao\IEEEauthorrefmark{2}}

\IEEEauthorblockA{\fontsize{10pt}{11pt}\selectfont \IEEEauthorrefmark{1}Huawei Munich Research Center, Germany \IEEEauthorrefmark{2}Technical University of Berlin, Germany \\
}

\IEEEauthorblockA{\IEEEauthorrefmark{3}Huawei Technologies Co., Ltd, China \IEEEauthorrefmark{4}CISUC, University of Coimbra, Portugal\\
}
\{qiao.yu, zhangwengui1, zhoumin27, yujialiang, shengzhenli, jorge.cardoso\}@huawei.com \\
\{jasmin.bogatinovski, odej.kao\}@tu-berlin.de \\
\vspace{-0.8cm}
}

\maketitle
\IEEEoverridecommandlockouts
\thispagestyle{plain}
\pagestyle{plain}

\begin{abstract}
Large-scale datacenters often experience memory failures, where Uncorrectable Errors (UEs) highlight critical malfunction in Dual Inline Memory Modules (DIMMs). Existing approaches primarily utilize Correctable Errors (CEs) to predict UEs, yet they typically neglect how these errors vary between different CPU architectures, especially in terms of Error Correction Code (ECC) applicability. In this paper, we investigate the correlation between CEs and UEs across different CPU architectures, including X86 and ARM. Our analysis identifies unique patterns of memory failure associated with each processor platform. Leveraging Machine Learning (ML) techniques on production datasets, we conduct the memory failure prediction in different processors' platforms, achieving up to 15\% improvements in F1-score compared to the existing algorithm. Finally, an MLOps (Machine Learning Operations) framework is provided to consistently improve the failure prediction in the production environment.
\end{abstract}


\section{Introduction}
With the expansion of cloud computing and big data services, the challenge of maintaining the Reliability, Availability, and Serviceability (RAS)\footnote{Reliability, Availability, and Serviceability are the foundational pillars used to measure the dependability of computer systems.} of servers is intensifying, due to memory failures, which represent a significant fraction of hardware malfunctions \cite{JD_HW,hw_failure,optical_failure}. These failures often occur as Correctable Errors (CEs) and Uncorrectable Errors (UEs). To tackle these issues, Error Correction Code (ECC) mechanisms such as SEC-DED \cite{SECDED}, Chipkill \cite{chipkill1997}, and SDDC \cite{intel_sddc} are utilized to detect and correct errors. For instance, Chipkill ECC is capable of correcting all erroneous bits from a single DRAM (Dynamic Random Access Memory) chip. However, its efficacy diminishes when errors span multiple chips, leading to system failures caused by UEs. Furthermore, the ECC mechanisms on modern Intel platform servers do not offer the same level of protection as Chipkill ECC, making them vulnerable to certain error patterns originating from a single chip \cite{Li_intel_bit_2022}. Therefore, relying exclusively on ECC mechanisms for memory reliability proves inadequate, as memory failures remain a prevalent source of system failures.

To enhance memory reliability, numerous studies \cite{Schroeder_DRAMErrors_2009,Meza_RevisitingError_2015,Sridharan_studyDRAM_2012,Sridharan_memory_error_modern_system,vilas_systematic_study,patel2022case_study,mattan_component_level_micro} have delved into the correlations between memory errors and failures, laying the groundwork for our research. Machine Learning (ML) techniques have been employed for predicting memory failures \cite{giurgiu_predicting_2017,du_memory_2018,du2020-intel,boixaderas_cost-aware_2020,Yu_drampakdd_2021,Du_nofreepredictor_2021,srds_in-Depth_MEM,fisrtCE_matter_TUB,alibaba_cuhk_uce_type,liustim_microsoft}, using CEs information from large-scale datacenters to forecast UEs. These investigations have effectively exploited the spatial distribution of CEs to improve memory failure prediction. Additionally, system-level workload metrics such as memory utilization, read/write access have been considered for memory failure prediction in \cite{sun2019-alibaba1,queen_workload-ware_2019,alibaba_workload-aware_2021}. Results from \cite{alibaba_workload-aware_2021} indicate that workload metrics play a minor role compared to other CE related features. Research in \cite{alibaba_2022_node_prediction} focuses on CE storms (a high frequency of CEs in a brief timeframe) and UEs to predict DRAM-caused node unavailability (DCNU), highlighting the significance of spatio-temporal features of CEs. Furthermore, \cite{Li_intel_bit_2022} explores specific error bit patterns and their association with DRAM UEs, developing rule-based indicators for DRAM failure prediction that vary by manufacturer and part number, adapted to the ECC designs of modern Intel Skylake and Cascade Lake servers. Moreover, \cite{huawei_2023_dsn,Yu_iccad_2023} examine the distribution of error bits and propose a hierarchical, system-level method for predicting memory failures, leveraging error bit characteristics. \textit{However, the incidence of UEs is influenced not just by DRAM faults but also by differences across CPU architectures, due to the diverse ECC mechanisms in use, which can alter the patterns of memory failures observed.} Understanding and modeling these failure patterns across various CPU platforms and ECC types is essential for accurate prediction of UEs. This gap in research motivates us to undertake the first study of DRAM failures comparing X86 and ARM systems, specifically the Intel Purley and Whitley platforms and the Huawei ARM K920 (anonymized to protect confidentiality) processor. By analyzing the relationship between UEs and fault patterns across these processor platforms, we aim to create targeted memory failure prediction algorithms. Additionally, we acknowledge the dynamic nature of server configurations, CPU architectures, memory types, and workloads. To address these variables, we introduce an MLOps framework designed to accommodate such changes, thereby continuously improving failure prediction throughout the lifecycle of the production environment.

We make the main contributions of this paper are below:
\begin{itemize}
    \item We present the first memory failure study between X86 and ARM systems, specifically focusing on Intel X86 Purley and Whitley, as well as Huawei ARM K920 processor platforms, in large-scale datacenters. Different fault modes within DRAM hierarchy are associated with memory failures across these platforms.
    \item We develop ML-based algorithms for predicting memory failures, leveraging identified DRAM fault modes to anticipate UEs on these platforms. 
    \item We establish an MLOps framework of failure prediction, to facilitate the collaboration across teams within the organization and help manage production ML algorithms lifecycle.
\end{itemize}

The organization of this paper is as follows: Section~\ref{sec:backgroundandmotivation} provides the background of this work. Section~\ref{sec:dataset} describes the dataset employed in our data analysis. Section~\ref{sec:Problem Formulation and performance-measures} details the problem formulation and performance metrics. Section~\ref{sec:fault_analysis} uncovers high-level fault modes within the DRAM hierarchy and their relationship to UEs across various platforms. Section~\ref{sec:failure_prediction} demonstrates the use of machine learning techniques for memory failure prediction. Section~\ref{sec: mlops_failure_prediction} introduces our MLOps framework for failure prediction. Related work is shown in Section~\ref{sec:related}. Section~\ref{sec:conclusion} concludes this paper.
\vspace{-0.2cm}
\section{Background}
\vspace{-0.1cm}
\label{sec:backgroundandmotivation}

\subsection{Terminology}
\label{ssec:terminology}
A \textit{fault} in DRAM acts as the root cause for an error, which may arise from a variety of sources, including particle impacts, cosmic rays, or manufacturing defects. 

An \textit{error} occurs when a DIMM sends incorrect data to the memory controller, deviating from what the ECC \cite{SECDED,chipkill1997,intel_sddc,supermicro_sddc} expects, indicative of an underlying fault. Memory errors, depending on the ECC's correction capacity, are classified into two main types: Correctable Errors (CEs) and Uncorrectable Errors (UEs). Two specific types of UEs are well described in \cite{giurgiu_predicting_2017}. 1) \textit{sudden UEs}, which result from component malfunctions that immediately corrupt data, and 2) \textit{predictable UEs}, which initially appear as CEs but evolve into UEs over time. Sudden UEs occur without prior CEs, whereas predictable UEs may be forecasted through CEs using algorithms designed for failure prediction. In this study, our focus is on predicting \textit{predictable UEs}, as they constitute the majority of memory failures, described in Section~\ref{sec:dataset}.

\vspace{-0.2cm}
\subsection{DRAM Organization and Access}
\label{sec:memory Access}
\begin{figure}[t]
\centering

\includegraphics[width=1\linewidth,keepaspectratio]{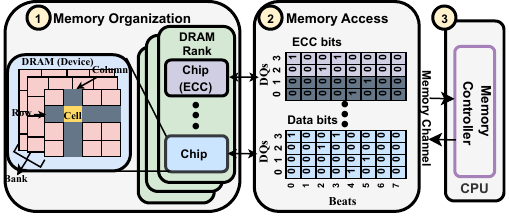}
\caption{Memory Organization.} 
\label{fig:mem_organization}
\vspace{-0.8cm}
\end{figure}

Fig.~\ref{fig:mem_organization} illustrates the memory's hierarchical layout and its CPU interactions. In Figure~\ref{fig:mem_organization}(1), it shows a DIMM rank made up of DRAM chips organized by banks, rows, and columns, where data moves from memory cells to the memory controller, which can generally detect and correct CEs. Figure~\ref{fig:mem_organization}(2) shows the data transmission of x4 DRAM DDR4 chips via Data Bus (DQs) upon CPU requests, involving 8 beats of 72 bits (64 data bits plus 8 ECC bits). Implementing the ECC, the memory controller detects and corrects them in Figure~\ref{fig:mem_organization}(3). Note that the exact ECC algorithms are highly confidential and never exposed and ECC checking bits addresses can be decoded to locate specific errors in DQs and beats. Finally, all these logs including Corrected and Uncorrected errors, events, and memory specifications are recorded in Baseboard Management Controller (BMC)\footnote{BMC is a specialized processor built into the server's motherboard, designed to supervise the physical status of computers, network servers, and additional hardware components.}. 

\subsection{Memory RAS Techniques }
\label{ssec:RAS}
DRAM subsystems leverage RAS features for protection, including proactive VM migrations to minimize interruptions and CE storm\footnote{\label{ce_storm}CE interruptions repeatedly occur multiple times, e.g., 10 times.} suppression to prevent service degradation. Advanced RAS techniques are designed to protect server-grade machines by avoiding faulty regions and employing sparing techniques like bit, row/column, and bank/chip sparing (e.g., Partial Cache Line Sparing (PCLS)\cite{Partial_cache_line_sparing}, Post Package Repair (PPR) \cite{spare_row_column}, Intel’s Adaptive Double Device Data Correction (ADDDC) \cite{du_page_offline_2021,spare_bank_chip}). Software-sparing mechanisms, such as the page offlining, mitigate memory errors\cite{du_page_offline_2021,Du_pageoffline_2019,mem_page_retirement}. However, these approaches may increase redundancy and overhead, affecting performance and limiting their universal applicability. Memory failure prediction plays a key role in foreseeing UEs and implementing specific mitigation strategies.

\vspace{-0.2cm}
\section{Dataset}
\label{sec:dataset}
Our dataset sourced from Huawei cloud datacenters, includes system configuration, Machine Check Exception (MCE) log, and memory events (CE storms, etc), focusing on DIMMs experiencing CEs and omitting those with sudden UEs due to the lack of predictive data. We examined error logs from approximately 250,000 servers across Intel Purley and Whitley platforms (including Skylake, Cascade Lake, and Icelake) as well as the Huawei K920 processor platform. 

Table~\ref{tab:data-description} describes an overview of our data, which includes over 90,000 DDR4 DIMMs from various manufacturers, spanning different CPU architectures, with CEs recorded from January to October 2023. Within Intel platforms, predictable UEs constitute 73\% of the UEs on the Purley platform, surpassing the rate of sudden UEs. In contrast, the Whitley platform shows a higher incidence of sudden UEs than Purley, despite it having a lower total UE rate compared to Purley. Meanwhile, in ARM system with K920 processor platform, there's a significant predominance of predictable UEs over sudden UEs, showcasing a variance in ratios compared to the X86 systems, the overall rate of UEs in the K920 dataset is less than that of the Intel platforms. Note that these statistics are specific to the datasets analyzed and the observed variations may be influenced by several factors, such as workload, server age, and distinct RAS mechanisms, etc. In particular, the ECC used in contemporary Intel platforms, which is integral for error correction and detection, is considered weaker than Chipkill. This is partly because some of the extra bits previously used by Intel ECCs are reallocated for other uses \cite{Li_intel_bit_2022}, such as to store ownership, security information, to mark failed areas of DRAM, etc. This suggests that the observed discrepancies in UE rates across different architectures could stem from the unique ECC mechanisms employed.
\vspace{-0.2cm}
\observation{The UE and sudden UE rates show variation between X86 and ARM systems. This discrepancy could be attributed to the distinct ECC mechanisms implemented within these differing architectures.}

\begin{table}[t]
    \centering
    \caption{Description of Dataset.}
    \begin{tabular}{|c|c|c|c|c|}
    \hline
          CPU    & DIMMs  & DIMMs  & Predictable UE   & Sudden UE     \\
           Platform &  with CEs& with UEs & DIMMs in \% & DIMMs in \%\\
          \hline
          Intel Purley   & $>$ 50,000 & $>$ 2,000 & 73\%  & 27\% \\
         \hline
          Intel Whitley    & $>$ 10,000  & $>$ 400 & 42\% & 58\%\\ 
         \hline
         K920   & $>$ 30,000  & $>$ 600 & 82\% & 18\%\\ 
         \hline
         
    \end{tabular}
    \label{tab:data-description}
    \vspace{-0.5cm}
\end{table}
\vspace{-0.2cm}
\section{Problem Formulation and Performance Measures}
\label{sec:Problem Formulation and performance-measures}
\begin{figure}[t]
\centering
\includegraphics[width=1\linewidth]{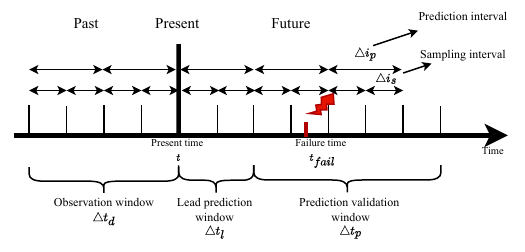}
\caption{Failure prediction problem definition \cite{huawei_2023_dsn}.} 
\label{fig:failure-prediction-problem}
\vspace{-0.7cm}
\end{figure}

The problem of predicting memory failures is approached as a binary classification task, following the methodology in \cite{huawei_2023_dsn,Yu_iccad_2023}. As illustrated in Figure~\ref{fig:failure-prediction-problem},  an algorithm at time $t$ uses data from a historical \textit{observation window} $\bigtriangleup t_{d}$ to predict failures within a future prediction period $[t+\bigtriangleup t_{l}, t + \bigtriangleup t_{l} + \bigtriangleup t_{p}]$, where $\bigtriangleup t_{l}$ represents the lead time \cite{time_machine_2023DSN} before a failure occurs, and $\bigtriangleup t_{p}$ is the duration of the prediction window. Event samples are recorded at intervals of $\bigtriangleup i_{s}$ (e.g., CE events are logged every minute), and predictions are made at intervals of $\bigtriangleup i_{p}$ (every 5 minutes). The observation ($\bigtriangleup t_{d}$) and \textit{prediction validation windows} ($\bigtriangleup t_{p}$) are set to 5 days and 30 days, respectively, to facilitate early proactive strategies. Note that these parameters were optimized based on empirical data from the production environment. The \textit{lead prediction time} $\bigtriangleup t_{l} \in (0, 3h]$, ranging up to 3 hours, designed to specific operational scenarios. A True Positive (TP) denotes an correctly predicted failure within the window. A False Positive (FP) represents an incorrect forecast. A False Negative (FN) describes a failure that happens without an earlier warning, and a True Negative (TN) is identified when no failures are anticipated or take place. The performance of the algorithm is evaluated using $Precision = \frac{TP}{TP + FP }$, $Recall = \frac{TP}{TP + FN }$ and $F1 = \frac{2\times Precision\times Recall}{Precision + Recall } $.

\begin{figure}[t]
\centering
\includegraphics[width=1\linewidth,keepaspectratio]{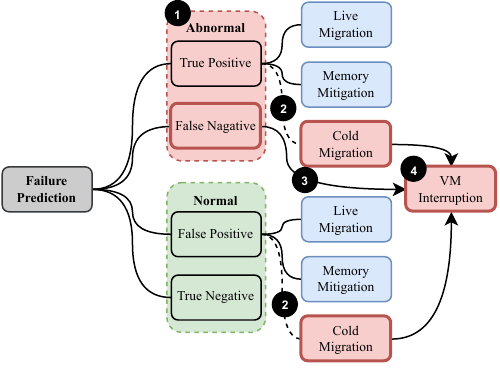}
\caption{VM Interruption under Failure Prediction.}
\label{fig:virr_evaluation}
\vspace{-0.3cm}
\end{figure}

\textbf{\textit{VM Interruption Reduction Rate (VIRR)}}. Prior works \cite{boixaderas_cost-aware_2020,Du_nofreepredictor_2021,alibaba_2022_node_prediction,Li_intel_bit_2022,huawei_2023_dsn,Yu_iccad_2023} have introduced cost-aware models to assess the benefits of memory failure prediction. In this work, we emphasizes \textit{VM Interruption Reduction Rate (VIRR)} \cite{huawei_2023_dsn} in Figure~\ref{fig:virr_evaluation}, as it more precisely reflects the effects on customer experience. 

Understanding the VIRR involves considering $V_{a}$ as the average number of VMs per server. Without predictive capabilities, the interruptions are calculated as \dingOne{} $V=V_{a}(TP + FN)$ as seen in the \textbf{Abnormal} of Figure~\ref{fig:virr_evaluation}. While proactive VM live migrations and memory mitigation techniques aim to minimize interruptions without service interruption, a notable fraction of VMs might still undergo cold migration, typically causing VM interruptions. Cold migrations are often the fallback when live migrations or memory mitigation are unfeasible, due to limited resources or unexpected failures, and are a common approach for VM reallocation and maintenance. The fraction of VMs under such migrations is denoted as $y_c$. As a result in Figure~\ref{fig:virr_evaluation}, we define \dingTwo{} $V'_{1} = V_{a} \cdot y_{c}(TP + FP)$ as the volume of interruptions from cold migrations triggered by accurate failure predictions (TP + FP). On the other side, missed failure predictions lead to increased interruptions, represented by \dingThree{} $V'_{2} = V_{a}\cdot FN$. Considering the prediction algorithm, the total interruptions are \dingFour{} $V' = V'_{1} + V'_{2}$. The VIRR formula is thus: $VIRR = \frac{V - V'}{V}$, simplifying to $(1 - \frac{y_{c}}{precision}) \cdot recall$, according to \cite{huawei_2023_dsn}.

In practical production settings, $y_{c}$ remains a positive value since VMs may need cold migrated due to the failure of live migration or memory mitigation. When a model's precision falls below the percentage of cold migration ($precision < y_{c}$), the VIRR shifts to negative, indicating an increase in VM interruptions. Conversely, high-precision models achieve positive VIRR, amplified by their recall rate. Based on our observations,  we've conservatively set $y_{c} = 0.1$ for our evaluation. Note that this value is already pessimistic, anticipating a reduction in $y_{c}$ as cloud infrastructure evolves and expands.

\section{Fault Analysis}
\label{sec:fault_analysis}
\begin{figure}[t]
\centering
\includegraphics[keepaspectratio]{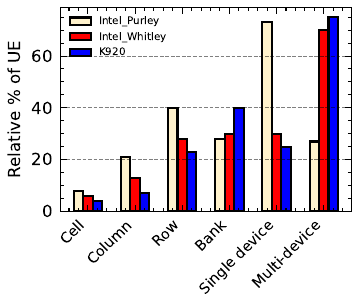}
\vspace{-0.3cm}
\caption{Relative \% of UE.} 
\label{fig:Relative_UE_percentage}
\vspace{-0.7cm}
\end{figure}

Our analysis investigates the high-level fault modes in the DRAM hierarchy, and correlates them with UE rates across various platforms. We consider faults at the DRAM-level, including cell, column, row, bank faults as illustrated in Figure~\ref{fig:mem_organization}(1). A \textbf{Cell fault} occurs when CEs in a cell surpass a set threshold, while \textbf{Row} and \textbf{Column faults} are identified by exceeding thresholds across a row or column, respectively. \textbf{Bank faults} arise when thresholds for both row and column faults within a bank are exceeded. Additionally, when CEs affect a single device, this constitutes a \textbf{Single-device fault}. In contrast, if CEs extend across multiple devices, it is \textbf{Multi-device fault}. Further details on fault definitions and threshold settings can be found in \cite{huawei_2023_dsn,Yu_iccad_2023,vilas_systematic_study}. The approach to calculating the relative UE rate depicted in Figure~\ref{fig:Relative_UE_percentage} follows previous studies \cite{Yu_iccad_2023,huawei_2023_dsn,Meza_RevisitingError_2015,Sridharan_memory_error_modern_system,srds_in-Depth_MEM}, categorizing DIMMs according to distinct fault types (e.g., cell faults) and assessing the percentage of DIMMs that encounter UEs.

As shown in Fig.~\ref{fig:Relative_UE_percentage}, the most UEs are attributed to faults in higher-level components, such as row and bank faults across all platforms. Specifically, on the Intel Purley platform, the primary source of UEs is single device faults. Conversely, in both the Whitley and K920 platforms, UEs predominantly arise from multi-device faults, possibly due to variations in ECC mechanisms.
\vspace{-0.2cm}
\observation{The Intel Purley platform primarily experiences UEs due to single device faults, a trend that appears to diminish in the Whitley platform. Meanwhile, the K920 platform exhibits fewer single device faults, potentially attributed to the efficiency of its K920-SDDC.}

\begin{figure*}[t]
\centering
\includegraphics[width=1\linewidth, keepaspectratio]{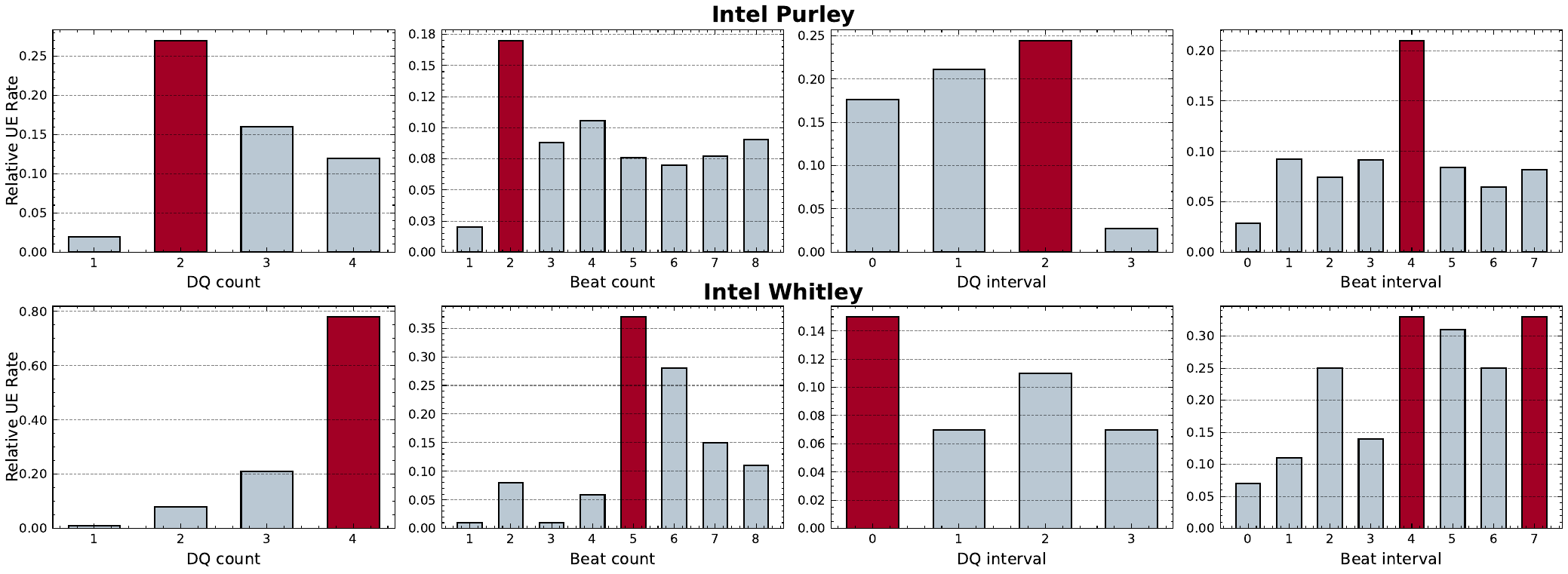}
\caption{Analyses of Error Bits in Intel Platforms: Highlighting The Highest Rate with Red Bar.} 
\label{fig:multi_platform_Error_bits_analysis}
\vspace{-0.3cm}
\end{figure*}

Then we investigate the failure patterns of error bits in DQs and beats, similar to \cite{Yu_iccad_2023}. As shown in Fig.~\ref{fig:multi_platform_Error_bits_analysis}, we examined the counts and intervals of error DQs and beats in x4 bit width DRAM. On the Intel Purley platform, 2 error DQs and beats counts and a 4-beat interval are associated with significantly higher UE rates, in comparison to other error DQ and beat counts and intervals. By contrast, the Intel Whitley platform exhibited higher UE rates with 4 error DQs and 5 error beats counts. However, variations in UE rates were not significantly influenced by the intervals between DQs and beats from our observations. Thus, the failure patterns on Intel's more advanced Whitley platform are markedly different from those observed on the Purley platform.
\vspace{-0.2cm}
\observation{At the bit-level within Intel architecture, distinct DQ and beat patterns emerge, highlighting the importance of formulating failure patterns designed to specific platforms due to the potential variations in their underlying ECC mechanisms.}

\section{Failure Prediction}
\label{sec:failure_prediction}

We develop our failure prediction using tree-based algorithms (Random Forest and
LightGBM \cite{Yu_iccad_2023,huawei_2023_dsn,alibaba_2022_node_prediction}) and deep learning FT-Transformer \cite{ft-transformer2021revisiting}. These ensemble learning techniques have been prevalently utilized in previous memory failure prediction literature \cite{alibaba_workload-aware_2021,Yu_drampakdd_2021,Du_nofreepredictor_2021,boixaderas_cost-aware_2020,giurgiu_predicting_2017}, with the FT-Transformer considered as the leading algorithm for handling tabular data in the field of deep learning. The experimental design and feature generation follow the methodology in \cite{Yu_iccad_2023}, which categorizes samples into two classes: Positive and Negative. DIMMs expected to experience at least one UE within the prediction window are categorized as \textbf{Positive}, while those expected not to have any UE are classified \textbf{Negative}. Samples labeling is based on the time interval between a CE and its subsequent UE, with specifics on interval settings available in \cite{Yu_iccad_2023,huawei_2023_dsn}. Features used in our models include DRAM characteristics such as manufacturer, data width, frequency, chip process, CE error rate, our conducted failure analysis, and memory events. The performance of these algorithms was evaluated using precision, recall, F1-score and VIRR as described in Section~\ref{sec:Problem Formulation and performance-measures}.

\begin{table*}[t]
\centering
\caption{Overview of Algorithm Performance Comparisons. (\textit{X} denotes  the absence of prediction values)}
\label{tab:Overview of Algorithm Performance Comparisons}
\vspace{-0.1cm}
\begin{tabular}{|c|cccc|cccc|cccc|}
\hline
\multirow{2}{*}{Algorithm} & \multicolumn{4}{c|}{Intel Purley} & \multicolumn{4}{c|}{Intel Whitley} & \multicolumn{4}{c|}{K920} \\ \cline{2-13} 
 & Precision & Recall & F1 & VIRR & Precision & Recall & F1 & VIRR & Precision & Recall & F1 & VIRR \\ \hline
Risky CE Pattern \cite{Li_intel_bit_2022} & 0.53 & 0.46 & 0.49 & 0.37 & X & X & X & X & X & X & X & X \\ 
\hline
Random forest & 0.61 & 0.62 & 0.61 & 0.52 & 0.34 & 0.46 & 0.39 & 0.32 & 0.44 & 0.51 & 0.47 & 0.39 \\ 
LightGBM & 0.54 & 0.80 & \textbf{0.64} & \textbf{0.65} & 0.46 & 0.54 & 0.49 & \textbf{0.45} & 0.51 & 0.57 & \textbf{0.54} & \textbf{0.46} \\ 
FT-Transformer & 0.49 & 0.74 & 0.59 & 0.58 & 0.53 & 0.49 & \textbf{0.50} & 0.40 & 0.40 & 0.54 & 0.46 & 0.41 \\ \hline
\end{tabular}
\vspace{-0.5cm}
\end{table*}

\textbf{Comparison with the existing approach.} We compare our algorithms with the existing reproduced Risky CE Pattern approach in \cite{Li_intel_bit_2022}, particularly for the Intel Skylake/Cascade (Purley platform) architecture. However, we noted a lack of dedicated memory failure prediction algorithms for the Intel Whitley platform and the Huawei ARM K920 Platform. Table~\ref{tab:Overview of Algorithm Performance Comparisons} shows the superior performance of our method, achieving the high F1-score of 0.64 and VIRR of 0.65 using LightGBM on the Intel Purley platform, outperforming rule-based risky CE pattern algorithm. Additionally, it scores 0.50 F1-score on the Intel Whitley platform using the FT-Transformer, and 0.54 F1-score and 0.46 VIRR in K920 architecture with LightGBM. The LightGBM results overall outperformed than other machine learning methods including deep learning FT-Transformer algorithm, which agrees with the finding in \cite{treethanDL_2022}. 
\vspace{-0.2cm}
\observation{Prediction efficacy varies across platforms; the Intel Whitley platform demonstrates comparatively weaker predictive performance than both the Intel Purley and K920 platform.}

\section{MLOps of Failure Prediction}
\label{sec: mlops_failure_prediction}
After developing machine learning (ML) algorithms that accurately predict memory failures, it becomes crucial to both maintain and enhance these algorithms and to automate their operation within the data center. The MLOps framework is ideally suited for this purpose, ensuring the continuous accuracy and applicable of our memory failure prediction algorithms. Figure~\ref{fig:mlops_framework} illustrates an overview of the MLOps framework for memory failure prediction, with the workflow introduced in stages as follows:

\begin{figure*}[t]
\centering
\includegraphics[width=1\linewidth,keepaspectratio]{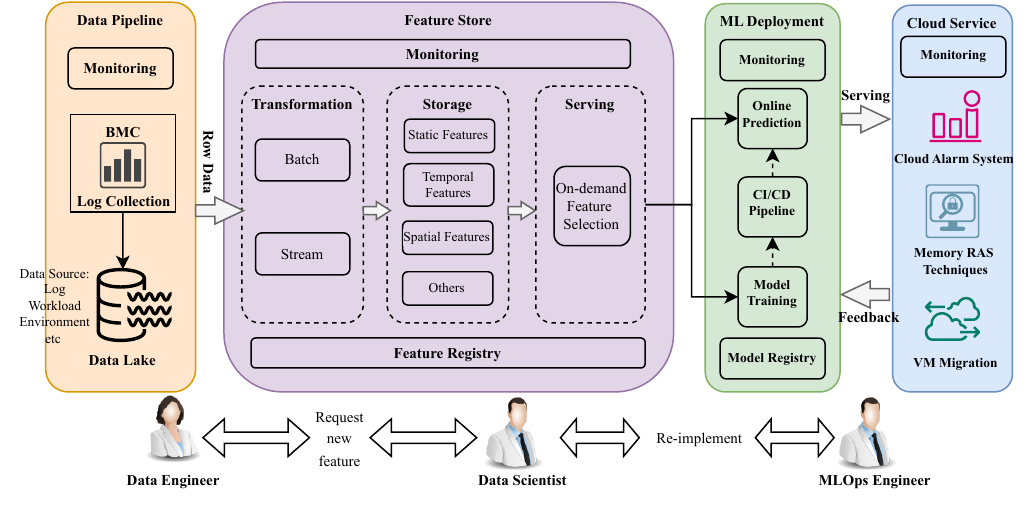}
\vspace{-0.8cm}
\caption{The MLOps Framework of Failure Prediction.} 
\label{fig:mlops_framework}
\vspace{-0.7cm}
\end{figure*}

\textbf{Data Pipeline}: The initial stage involves collecting raw data from various sources. For example, CE and UE logs are collected by the BMC, and are then processed and stored in the Data Lake, alongside other data sources such as runtime workload metrics (e.g., CPU utilizations) and environmental metrics (e.g., server locations, temperatures) using the Huawei Data Lake Insight (DLI) solution.

\textbf{Feature Store}: A feature store acts as a centralized repository for transforming, storing, cataloging, and serving features used in model training and inference. It ensures consistency between training and prediction phases and accelerate the development of machine learning models by making features readily accessible. This involves:
\begin{itemize}
    \item \textbf{Transformation}: Converting raw data into features suitable for machine learning algorithms. This process is divided into batch and stream transformations for model training and online prediction, respectively.
    \item \textbf{Storage}: Once transformed, features are stored in an accessible format for training and prediction. They are cataloged with registry information to standardize features across all teams’ models. For example, CEs are converted into temporal and spatial features within the DRAM hierarchy. This conversion includes the distribution of error bits across DQs and beats, the number of faults, within different time intervals (1min, 1h, 5d, etc) and memory configurations, such as manufacturers, DRAM processes are further encoded to static features.
    \item \textbf{Serving}: Feature store serves features for model training and inference, enabling Data Scientists to select features on demand for training different models based on varying requirements. For instance, Data Scientists might develop various models designed to distinct CPU architectures, utilizing unique features for each.
\end{itemize}

\textbf{ML Deployment}: This phase involves (1) \textbf{model training}, which contains algorithms selection, hyperparameters tuning, and the application of these configured algorithms on prepared datasets. This task can be performed manually by Data Scientists or through automated tools like AutoML. Once models are trained, and show substantial improvements in predefined benchmark evaluations, they advance to deployment in the production environment. This deployment leverages a (2) \textbf{Continuous Integration and Continuous Delivery (CI/CD) pipeline}, which automates the integration, testing, and deployment of ML algorithms, thereby ensuring models can be consistently updated and reliably released within the production environment. Subsequently, the successfully deployed models continue delivering (3) \textbf{online prediction} utilizing streaming data and the prediction results will be served to our cloud service.

\textbf{Cloud Service}: The alarm is triggered in the \textbf{Cloud Alarm System} upon predicting memory failures. Depending on different use cases, the \textbf{memory RAS techniques} are then implemented to mitigate the failures, with VMs being migrated automatically or manually as required. 

\textbf{Monitoring}: Throughout the MLOps workflow, each phase is continuously monitored through dashboards. This includes monitoring data collection rates, feature importance, and algorithm performance, as well as VM migrations and service interruptions. To enhance algorithm accuracy and ensure fairness in predictions, feedback is proactively gathered from the cloud service. Dashboards are implemented in both testing and production settings to closely observe algorithm performance, as well as the rates of false negatives and positives. This dual-environment monitoring facilitates the ongoing refinement of algorithms.

In our memory failure prediction development, collaboration across various teams is essential, this collaborative effort spans from Data Engineers, who are responsible for processing new data and integrating it into the Data Lake in response to Data Scientists' requests. Data Scientists analyze this data, develop predictive algorithms, and specify requirements for operational deployment. To the MLOps Engineers, who take on the critical role of re-implementing and deploying newly developed algorithms by Data Scientists into the production environment.

\vspace{-0.4cm}
\section{Related Work}
\label{sec:related}
Empirical research \cite{Meza_RevisitingError_2015,Schroeder_DRAMErrors_2009,Sridharan_studyDRAM_2012,Sridharan_memory_error_modern_system,vilas_systematic_study} has laid the groundwork in the study of memory errors, focusing on correlation analyses and failure modes. Their works serves as foundational elements for developing memory failure prediction algorithm. This section highlights significant contributions in memory failure prediction.

The ensemble learning approaches~\cite{giurgiu_predicting_2017,du2020-intel,Du_nofreepredictor_2021,boixaderas_cost-aware_2020} have constantly improved memory failure prediction by leveraging correctable errors, event logs, sensor metrics. The node/server-level memory unavailability prediction methods are proposed in ~\cite{alibaba_2022_node_prediction,srds_in-Depth_MEM}, considering both UE and CE storm/CE-driven prediction. Li et al.~\cite{Li_intel_bit_2022} explored correlations between CEs and UEs using error bit information and DIMM part numbers, creating a new risky CE indicator for UE prediction across different manufacturers and part numbers. Peng et al.~\cite{alibaba_cuhk_uce_type} designed DRAM failure forecasters by utilizing different UCE types.
Yu et al.~\cite{huawei_2023_dsn,Yu_iccad_2023} further examined the distribution of error bits and proposed a hierarchical, system-level approach for predicting memory failures by utilizing the error bits features.

However, the literature mentioned does not examine failure patterns across various processors' platforms, nor does it engage in the development of ML models specifically designed for distinct CPU architectures to improve prediction. In our previous work \cite{Yu_icdcs_2024}, we explored memory failure patterns across various CPU architectures. In this extended version, we further expand the fault analysis, present 4 findings, and establish an MLOps framework to continuously improve failure prediction models throughout their lifecycles.
\vspace{-0.2cm}

\section{Conclusion}
\label{sec:conclusion}
\vspace{-0.1cm}
We present the first comprehensive analysis of DRAM failures spanning both X86 and ARM systems across various platforms in large-scale datacenters. From our analytical and predictive modeling work, we report 4 findings: 1) UE and sudden UE rates differ between X86 and ARM systems. 2) Fault modes vary across architectures. 3) Bit-level failure patterns of DRAM are architecture-dependent. 4) Prediction accuracy differs by platforms. Utilizing datasets from production environment, our approach showcased a 15\% enhancement in F1-score compared to the method in \cite{Li_intel_bit_2022}, specifically within the Intel Purley platform. Moreover, we executed initial experiments on the Intel Whitley and ARM-based platforms, achieving F1-scores of 0.50 and 0.54, along with VIRR of 0.45 and 0.46 respectively. Finally, we present our MLOps framework for memory failure prediction, implemented in the production environment, This framework is designed to ensure the continuous enhancement and maintenance of failure prediction performance.
\vspace{-0.1cm}
\section*{Acknowledgement}
\vspace{-0.1cm}
We thank the anonymous reviewers from DSN'24 for their constructive comments.

\bibliographystyle{IEEEtran}
\bibliography{bibliography}

\end{document}